\documentclass[aps,prl,twocolumn,groupedaddress]{revtex4-1}
\usepackage{graphicx}% Include figure files
\usepackage{dcolumn}% Align table columns on decimal point
\usepackage{bm}% bold math
\usepackage[normalem]{ulem} 

\usepackage{soul}
\usepackage[dvipsnames]{xcolor}
\usepackage{float}

\begin{document}
	
% highlight changes by person	
\newcommand{\gc}[1]{\textcolor{red}{#1}}
\newcommand{\bh}[1]{\textcolor{blue}{#1}}
\newcommand\del{\textcolor{darkgray}\bgroup\markoverwith
	{\textcolor{red}{\rule[0.5ex]{2pt}{0.8pt}}}\ULon}

\newcommand{\cmnt}[1]{\textcolor{orange}{#1}}
\newcommand{\ignore}[1]{}

% highlight changes by person

%\preprint{}

\title{Exceeding the asymptotic limit of polymer drag reduction}


\author{George H. Choueiri}
%\thanks{}
\affiliation{Institute of Science and Technology Austria, 3400 Klosterneuburg, Austria}

\author{Jose M. Lopez}
%\thanks{}
\affiliation{Institute of Science and Technology Austria, 3400 Klosterneuburg, Austria}

\author{Bj\"{o}rn Hof}
%\thanks{}
\affiliation{Institute of Science and Technology Austria, 3400 Klosterneuburg, Austria}


\date{\today}

\begin{abstract}
The drag of turbulent flows can be drastically decreased by addition of small amounts of high molecular weight polymers. While drag reduction initially increases with polymer concentration, it eventually saturates to what is known as the maximum drag reduction (MDR) asymptote; this asymptote is generally attributed to the dynamics being reduced to a marginal yet persistent state of subdued turbulent motion.  Contrary to this accepted view we will show in the following that for an appropriate choice of parameters polymers can reduce the drag beyond the suggested asymptotic limit, eliminating turbulence and giving way to laminar flow. At higher polymer concentrations however, the laminar state becomes unstable, resulting in a fluctuating flow with the characteristic drag of the MDR asymptote. Our findings indicate that the asymptotic state is hence dynamically disconnected from ordinary turbulence.
\end{abstract}

%\keywords{}
\maketitle

In pipe and channel flows turbulence is often responsible for more than 90\% of the friction losses. A very efficient and often used method to reduce this frictional drag is by addition of small amounts of long chain polymers. Since its discovery nearly 70 years ago this effect has been studied extensively and various theories have been put forward to explain the mechanism of drag reduction (DR). It is commonly found that DR increases with polymer concentration but eventually saturates to the ``maximum drag reduction'' (MDR) asymptote \cite{virk1970ultimate}, as shown in Fig.~\ref{fig:friction_factor}$a$. All studies to date suggest that regardless of polymer concentration this limit cannot be overcome \cite{sreenivasan2000onset,white2008mechanics,xi2012dynamics}. The MDR asymptote is found to be identical for different types of polymers and polymer solvent combinations \cite{virk1970ultimate}. Based on these observations the general view is that polymers decrease turbulent activity (either via elastic \cite{tabor1986cascade} or viscous effects \cite{lumley1969drag} or both) and that eventually turbulence is reduced to a marginal state which corresponds to the MDR asymptote. More explicitly it has been argued \cite{procaccia2008colloquium,xi2012dynamics} that turbulence is minimized to the edge between laminar and turbulent motion. It is however not clear how turbulence can persist in this limit. In purely Newtonian flow the edge is intrinsically unstable and separates initial conditions that go turbulent from those that collapse back to laminar and it has not been shown if and how this state would become stable due to the action of polymers.\\
 An alternative interpretation of the MDR state has been given more recently \cite{samanta2013elasto}, where it has been observed that with increasing polymer concentration a different instability sets in; due to its occurrence at finite inertia and the elastic nature of the polymers it has been dubbed elasto-inertial instability. This instability has been observed independently in direct numerical simulations \cite{dubief2013direct} and in laboratory experiments \cite{samanta2013elasto}. In the experimental study it could be shown that the elasto-inertial instability strictly arises at Reynolds numbers below those at which the MDR asymptote is assumed and that at the same time the transition to turbulence is delayed.  Also it has been observed (in agreement with earlier studies \cite{ram1964structural,little1970drag} that for moderately high polymer concentrations, chaotic motion sets in at Reynolds numbers ($\text{Re} = U_bD/\nu$) $\approx 900$  (see also \cite{ram1964structural,little1970drag}), much below those at which turbulence can be observed in Newtonian fluids ($\text{Re} \approx 2000$). Here $U_b$ is the bulk flow velocity, $D$ is the pipe diameter, and $\nu$ is the kinematic viscosity of the fluid. Based on these observations it has been proposed \cite{samanta2013elasto} that on MDR the dynamics are driven by the elasto-inertial instability while Newtonian turbulence (NT) is eliminated before reaching MDR and the authors dubbed the corresponding dynamical state elasto-inertial turbulence (EIT). \\
In the following we will demonstrate that polymers can for an appropriate choice of parameters eliminate fully turbulent motion. For increasing polymer concentration the laminar flow eventually becomes unstable again giving rise to the MDR state, which is hence disconnected from NT. As will be shown, flows in the asymptotic drag limit at high Reynolds numbers structurally differ from NT and at the same time closely agree with the characteristic streak patterns resulting from EIT at Reynolds numbers well below the threshold for NT.\\
\begin{figure*}
	\includegraphics[width=17.2cm]{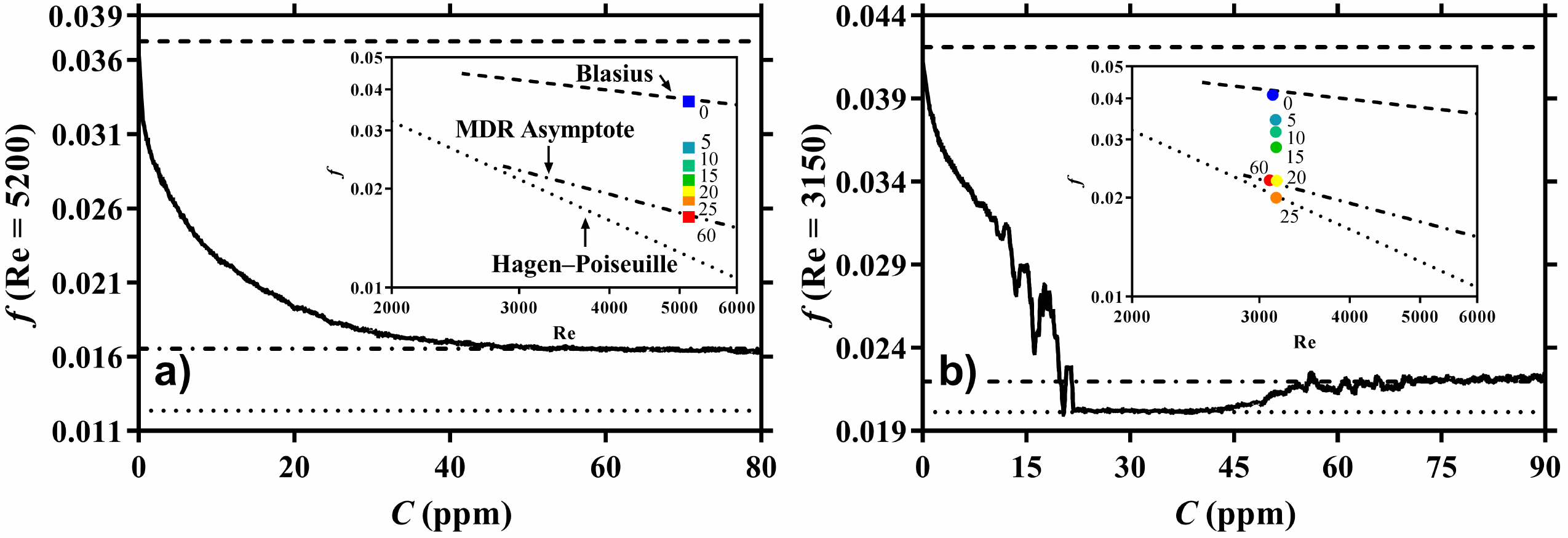}% Here is how to import EPS art
	\caption{\label{fig:friction_factor} The main figures show the Darcy--Weisbach friction factor as a function of continuously increasing polymer concentration while the insets show the friction factor at distinct concentrations constant Reynolds number. Dashed lines - Blasius correlation ($f=0.316\text{Re}^{-1/4}$), Dotted lines - Hagen--Poiseuille flow ($f=64/\text{Re}$), Dash--Dot lines - MDR Asymptote ($f=2.36\text{Re}^{-0.58}$).}
\end{figure*}
Experiments are carried out for pipe flow with water as the base fluid. A concentrated polymer solution was injected into the water at the pipe entrance. Dye was added to the concentrated polymer solution to verify that the fluid was uniformly mixed and homogeneous well upstream of the first measurement station located $480D$ downstream (see supplement for details). In the first set of experiments the Reynolds number was held fixed at 5200 starting from Newtonian flow where the friction factor is found to agree with the Blasius correlation for friction factors. When the concentration, $C$, is increased in distinct steps from 0 parts per million ($\text{ppm}$), by weight, to a maximum of $60\text{ppm}$ the friction factor is observed to monotonically decrease and eventually settle on the maximum drag reduction asymptote (Fig.~\ref{fig:friction_factor}$a-\text{inset}$). Note that the addition of polymers causes a viscosity increase in the fluid which has been measured and is taken into account for the quoted Reynolds number values for all the measurements reported.  When the experiment is repeated and $C$ is increased linearly from 0 to $80\text{ppm}$ (over a time span of $\approx 60,000$ advective time units in a quasi-steady fashion -- Fig.~\ref{fig:friction_factor}$a$), the continuous decrease of the friction factor and the monotonic approach towards MDR ($\sim48\text{ppm}$) are clearly seen. This observation precisely complies with the standard picture of polymer drag reduction and its asymptotic limit (see Figure 3 in \cite{white2008mechanics}).\\
A very different scenario is observed when Re is set to 3150 (Fig.~\ref{fig:friction_factor}$b-\text{inset}$). Again beginning on the Blasius line for NT an increase in polymer concentration to $20\text{ppm}$ appears to reduce the friction factor on average to the predicted MDR value. However, a further increase to $25\text{ppm}$ pushes the friction factor below what was believed to be the limiting threshold in polymer drag reduction and recovers the laminar (Hagen-Poiseuille) value. In addition measured fluctuation levels drop to the level of instrument noise recorded at zero flow. A further increase in polymer concentration surprisingly destabilizes the laminar flow: fluctuations increase and the friction factor increases to the ``maximum'' drag reduction asymptote ($60\text{ppm}$). While the average friction factors for 20 and $60\text{ppm}$ are almost identical and comply with the MDR value, structurally the flows are very different (see Fig.~\ref{fig:PIV_Re3150}$b$ \text{v.s.} $e,f$ where red and blue marked regions that strongly deviate from the mean streamwise speed $\text{i.e.}$ streaks). At $20\text{ppm}$ the flow is intermittent consisting of localized bursts of activity separated by much more quiescent regions. For $ C > 50\text{ppm}$ however the entire flow is in a fluctuating state (Fig.~\ref{fig:PIV_Re3150}$e,f$), but notably fluctuation levels are much reduced when compared to the bursts at $20\text{ppm}$. This picture is confirmed when the concentration is increased continuously at a slow rate (from 0 to $90\text{ppm}$ over the course of $\approx 60,000$ advective time units -- Fig.~\ref{fig:friction_factor}$b$) while Re is held fixed at  $\text{Re}=3150$. The friction factor decreases gradually and close to $20\text{ppm}$ drops more steeply to the laminar flow value. Laminar flow persists up to about $40\text{ppm}$ and then the friction factor begins to increase until it settles to the MDR value for $\text{C} > 55\text{ppm}$.\\
These experiments were repeated following two alternative protocols to ensure that the results are robust and independent of the detailed procedure. In the first case polymers were injected $~150D$ further downstream into the fully turbulent flow and the results obtained were identical to the above within experimental uncertainties. In the second case, as opposed to polymer injections, experiments were carried out with premixed polymer solutions of set concentrations. Equally in this case a laminar window was found for $Re \lesssim 3600$ separating drag reduced turbulence at lower concentration and MDR at higher (see supplement for more details).\\
\begin{figure}
	\includegraphics[width=8.6cm]{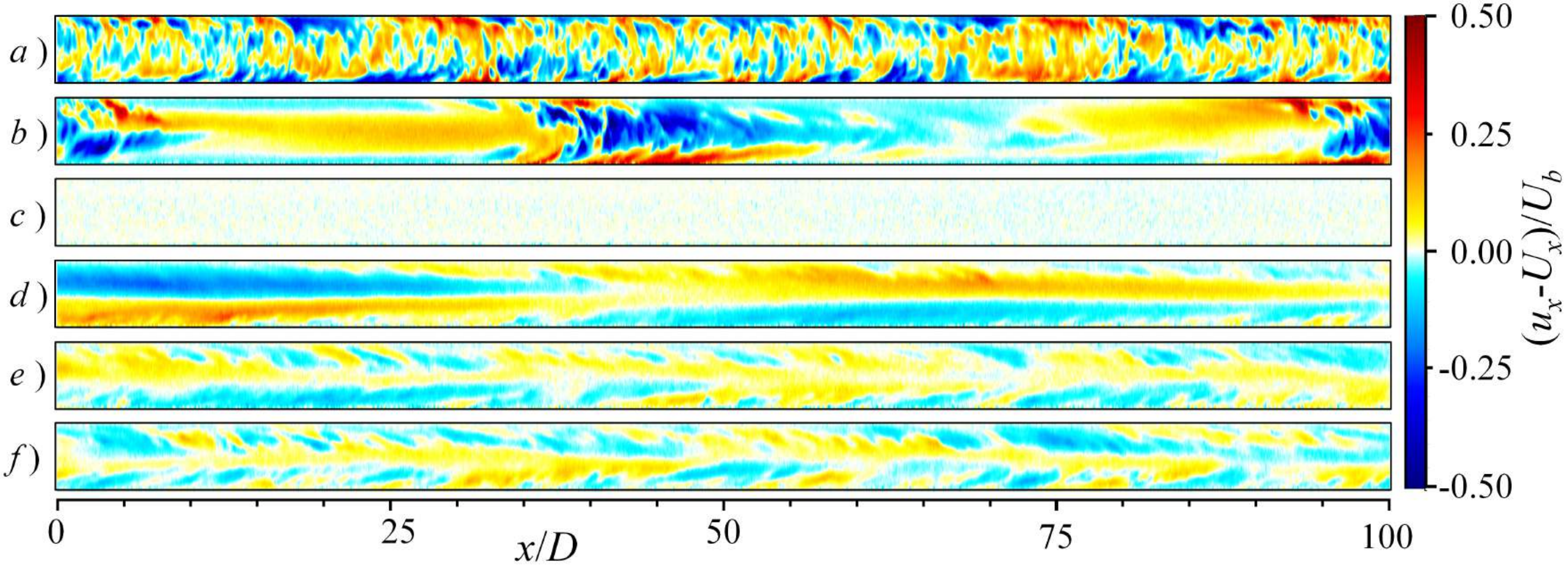}% Here is how to import EPS art
	\caption{\label{fig:PIV_Re3150} Streamwise velocity deviations with respect to the mean flow profile. High speed streaks appear in red, low speed streaks in blue. The velocity field was obtained from PIV images taken in a cross-sectional plane $\sim3D$ in length. The larger structures are then reconstructed by assuming Taylor's frozen turbulence hypothesis; i.e. assuming that turbulence is advected downstream quickly and changes in time are slow. Images taken at different times are assembled and matched to regain the spatial structure. Measurements are carried out at $\text{Re}=3150$ and polymer concentrations of: a)~$0\text{ppm}$	b)~$20\text{ppm}$ c)~$30\text{ppm}$ d)~$50\text{ppm}$ e)~$100\text{ppm}$ and f)~$150\text{ppm}$. Figure scaled to 20\% in the horizontal direction. $u_x$ is the local instantaneous streamwise velocity, $U_x$ the average streamwise velocity and $U_b$ the bulk velocity. The flow direction is left to right.}
\end{figure}
To elucidate the qualitative difference in the drag reduction scenario observed at the two different Reynolds numbers presented in Fig.~\ref{fig:friction_factor} we carry out a detailed investigation of the Reynolds number -- polymer concentration parameter space. In particular we observe that the onset of turbulence is delayed (left lower branch in Fig.~\ref{fig:stability_map}) by the action of polymers. While in previous studies \cite{samanta2013elasto} the delay only extended to parameters where in the Newtonian case flows at most are spatio-temporally intermittent ($\text{Re}\approx2600$) in the present case the delay extends to $\sim 25\%$ larger Reynolds numbers where in the Newtonian case turbulence is space filling ($\text{Re} \gtrsim 2800$) and assumes its characteristic friction factor scaling (Blasius). For increasing Re in this regime ($C\lesssim20\text{ppm}$ in Fig.~\ref{fig:stability_map}) turbulence sets in in the form of localized turbulent structures (puffs) and subsequently to growing turbulent structures, commonly referred to as slugs. Like the onset of puffs the onset of slugs is equally delayed compared to Newtonian fluids. In contrast at higher concentrations ($\gtrsim30\text{ppm}$) a qualitatively different instability is encountered. Here fluctuations set in more uniformly in space  (Fig.~\ref{fig:PIV_Re3150}$d$) lacking the spatial intermittent character of the Newtonian transition scenario. This observation is in qualitative agreement with the previous study on elasto-inertial instability \cite{samanta2013elasto}. For a further increase in concentration ($C \gtrsim 90 ppm$) this instability occurs at Re distinctly below the lowest Re where NT would be observed.\\
\begin{figure}
	\includegraphics[width=8.6cm]{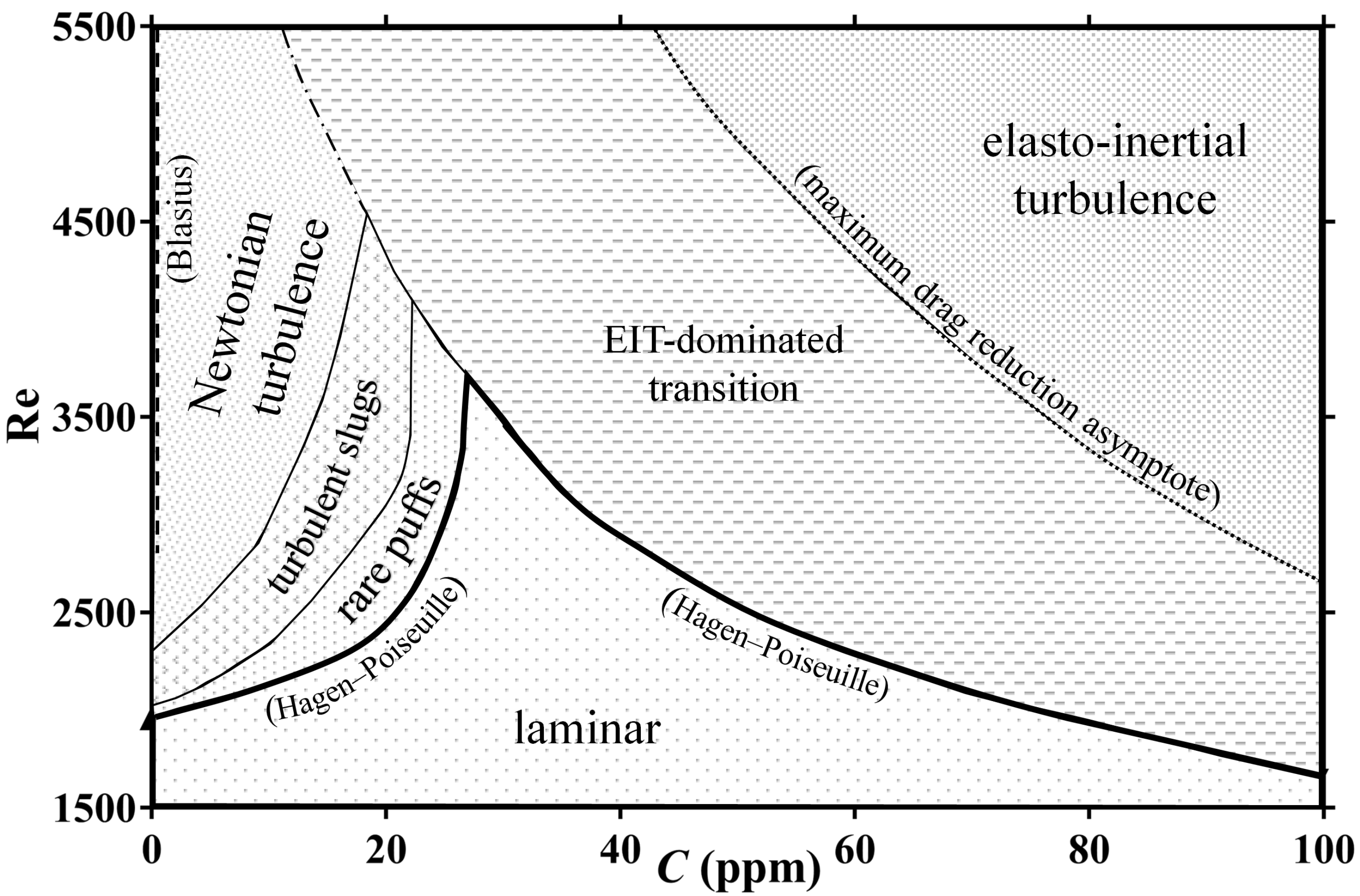}% Here is how to import EPS art
	\caption{\label{fig:stability_map}State map for Reynolds number versus polymer concentrations.}
\end{figure}
Starting from Newtonian turbulence ({$C=0$) for a fixed Re (Fig.~\ref{fig:stability_map} -- Blasius line on the left) and increasing polymer concentration, if $\text{Re} \leq 4500$ fully turbulent motion becomes unstable and returns to a regime of localized turbulent patches interspersed by more quiescent regions. Upon further increase in concentration (for $\text{Re} \lesssim 3600$) the localized turbulent structures are found to collapse and the flow completely relaminarizes. This scenario is the inverse of the familiar turbulence transition scenario in Newtonian pipe flow where turbulence first appears in the form of localized patches (first puffs then slugs) and only upon further increase in Re fully turbulent flow becomes stable \cite{barkley2015rise}. For even larger polymer concentration the laminar flow becomes unstable as ``elasto-inertial'' instability sets in (right lower branch in Fig.~\ref{fig:stability_map}). Note that here the complete suppression of NT takes place before the polymer driven elasto-inertial instability occurs. At somewhat larger Reynolds numbers ($3600<\text{Re}<4500$) turbulence breaks up into localized puffs / slugs, however before complete relaminarization is observed the instability to EIT occurs resulting in a mixed state which then eventually approaches MDR. At higher Re ($\text{Re}>4500$) the scenario seemingly follows the traditional view of MDR where NT is continuously suppressed by the action of polymers and the friction factor eventually settles to MDR. However what has been overlooked in previous studies is that at intermediate concentrations elasto-inertial instability sets in before MDR is reached.\\
\begin{figure}
	\includegraphics[width=8.6cm]{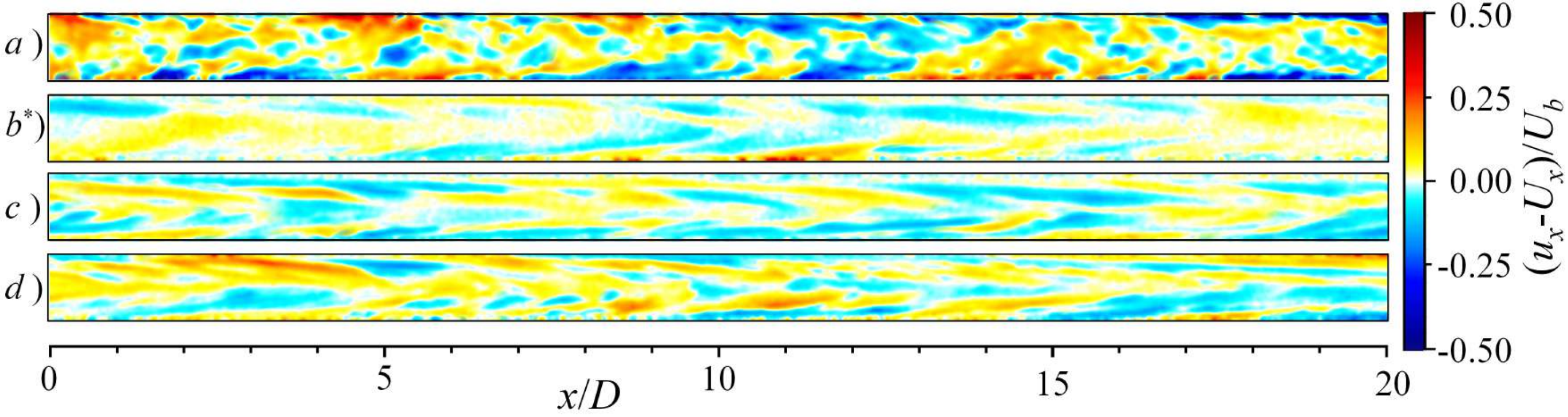}% Here is how to import EPS art
	\caption{\label{fig:stitched_images}  Streamwise velocity deviations with respect to the mean flow profile for various Reynolds numbers and polymer concentrations. a)~$\text{Re}=5200$ ($0\text{ppm}$)  b*)~$\text{Re}=1000$ ($70\text{ppm}$ – 50\% glycerol) c)~$\text{Re}=5200$ ($125\text{ppm}$)  d)~$\text{Re}=10000$ ($68\text{ppm}$).}
\end{figure}
The reverse transition from fully turbulent flow to localized  slugs and puffs (for $\text{Re}<4500$) and eventually laminar flow (for $\text{Re}<3400$) and the fact that onset of the elasto-inertial instability is always encountered prior to the final approach to MDR raise the question if also at larger Re NT is eventually fully suppressed as the polymer concentration is increased. In order to elucidate this question we will in the following compare the flow structure in the MDR limit for different Reynolds numbers. When comparing flow structures between Newtonian (Fig.~\ref{fig:PIV_Re3150}$a$) and MDR (Fig.~\ref{fig:PIV_Re3150}$f$) for flow at $\text{Re}=3150$, it is apparent that the streak pattern characteristic for NT is virtually absent in the MDR flow. In the latter case only weak, much more elongated streaks that are slightly inclined with respect to the flow direction are present. As noted above, these structures arise after NT has been eliminated and are a consequence of the elasto-inertial instability, directly arising from a state of laminar flow. The MDR flow at $\text{Re}=5200$ and $10000$ (Fig.~\ref{fig:stitched_images}$c,d$) qualitatively resembles that at $\text{Re}=3150$ (Fig.~\ref{fig:PIV_Re3150}$e,f$.) Also here the characteristic streak patterns of NT have disappeared and only weak, elongated streaks are visible. Overall the resemblance of flow structures on MDR for the three Reynolds numbers show that EIT has characteristic features that are independent of Re. To further illustrate this point, we increased the solution viscosity using a 50\% glycerol solution as to trigger the elasto-inertial instability at Re below that which can sustain NT ($\text{i.e.}$ 1000 in this case -- Fig.~\ref{fig:stitched_images}$b^*$) and we again observe low amplitude, elongated streaks similar to those on MDR at higher Re. Since the flows at $\text{Re}=1000$ and $3150$ are clearly disconnected and distinct from NT and result solely from the elasto-inertial instability, we propose that also at $\text{Re}=5200$ and $10000$ (where MDR is approached in the usual manner) NT is eventually marginalized, if not fully suppressed, and replaced by EIT only that here the eventual state is preceded by a co-existence phase rather than by relaminarization. Finally we computed the Reynolds stresses $\overline{u'v'}$, where $u'$ and $v'$ are the velocity fluctuations in the streamwise and radial direction respectively. As shown in Fig.~\ref{fig:reynolds_stress} flows in the MDR/EIT limit have markedly lower Reynolds stresses than their Newtonian counterparts (in agreement with earlier Reynolds stress measurements at MDR \cite{warholic1999influence}). We argue that the much lower Reynolds stress level further distinguishes flows in the MDR limit as a separate dynamical state.\\
\begin{figure}
	\includegraphics[width=8.6cm]{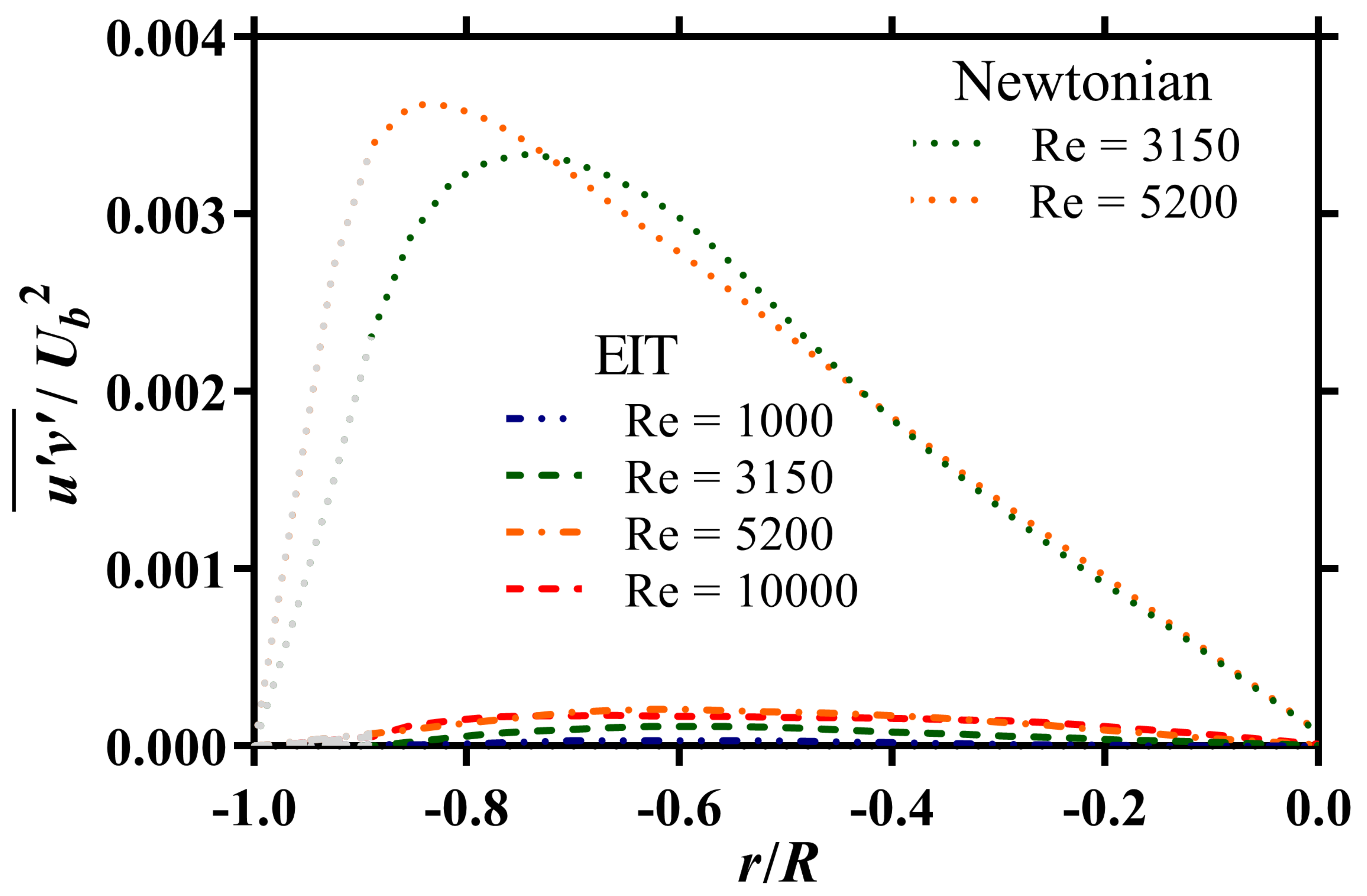}% Here is how to import EPS art
	\caption{\label{fig:reynolds_stress} Reynolds stresses for NT and EIT normalized by the square of the bulk flow velocity as a function of the radial position $r$, normalized by the pipe radius $R$. Due to higher uncertainty in near-wall PIV measurements, results in this region were forced to zero at the wall and are marked in gray were this was done.}
\end{figure}
In summary we have shown that polymer additives can fully eliminate NT and give rise to laminar flow. In the parameter regime  where this occurs a further increase in polymer concentration gives rise to an elasto-inertial instability \cite{samanta2013elasto} which leads to a drag increase approaching the MDR asymptote. In this case MDR does not correspond to the maximum drag reduction achievable by polymers. We propose that at larger Reynolds numbers ($\text{i.e.}$ larger shear rates) it is this elasto-inertial instability which keeps flows from relaminarizing and that this is ultimately responsible for the MDR state. Visualizations of the velocity fields of elasto-inertial turbulence at low Re, of MDR after relaminarization and of MDR at high Re feature very similar large scale structures and these substantially differ from the structure of ordinary Newtonian turbulence. Finally Reynolds stresses of EIT both at $\text{Re}=1000$ and higher Re MDR flows are at a similar level and distinct from Reynolds stress levels in Newtonian turbulence. An aspect of interest for applications is that the relaminarized flow at $\text{Re}<3600$ is not only stable to infinitesimal perturbations but also finite amplitude perturbations tend to swiftly decay. Hence here the full drag reduction potential is achieved and remains robust to disturbances. In larger pipe diameters where shear rates are lower the elasto-inertial instability is expected to appear only at larger Re, consequently the relaminarization interval prior to MDR may shift to larger Reynolds numbers where the drag reduction would be considerably larger.\\
\bibliography{Choueiri_Hof_2017_GC11}
\section{Supplemental Material}
\begin{figure*}
	\includegraphics[width=12.9cm]{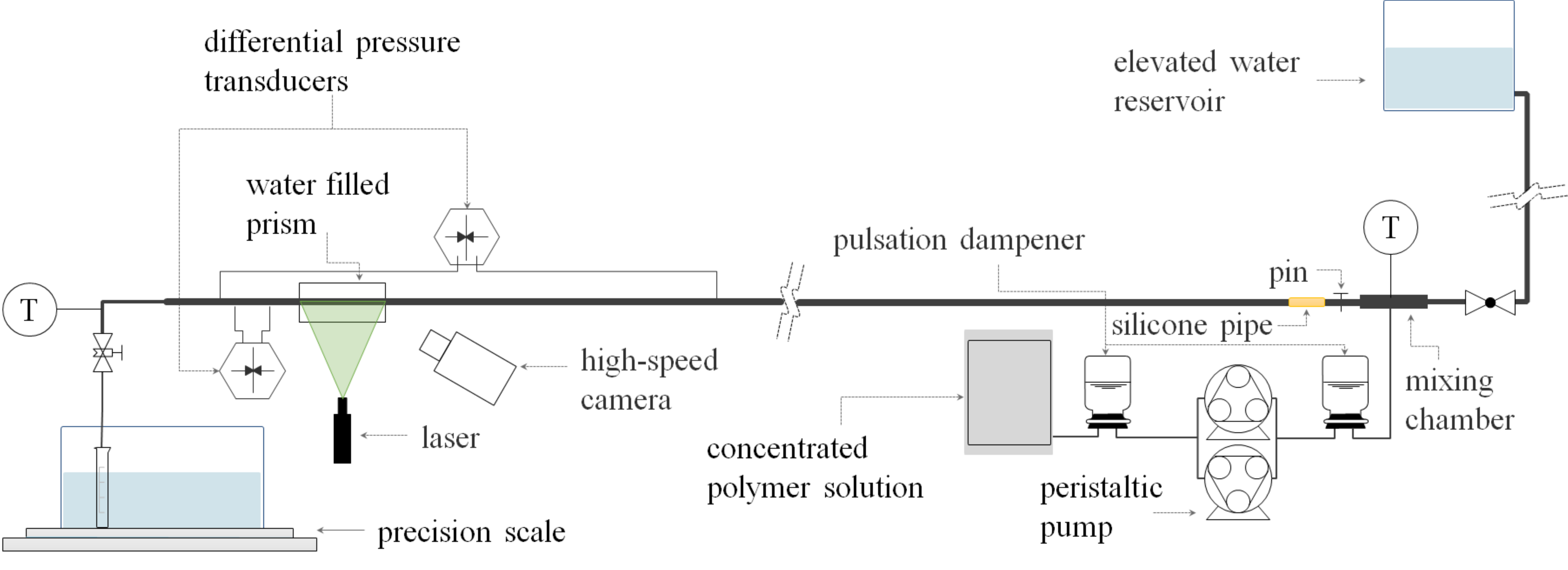}% Here is how to import EPS art
	\caption{\label{fig:sketch}Sketch of experimental setup; not to scale.}
\end{figure*}
Experiments are carried out in a $D=10\text{mm}$ glass pipe (where $D$ is the pipe diameter) with a total length of $646D$ using water at 296K as base for the working fluid (see Fig.~\ref{fig:sketch}). The main flow is fed from an elevated water reservoir of adjustable height (generally set between 2-3m above the outlet of the setup) which combined with a needle valve preceding the outlet allowed for precise flow rate control. Flow enters the pipe from an enlarged mixing chamber through a flush edge entrance and is further perturbed by a span-length 2mm pin $2.3D$ downstream from the inlet. In this configuration, Newtonian flow becomes fully turbulent for $\text{Re}>2800$. For lower Re ($1900<\text{Re}<2800$) the flow is dominated by spatio-temporal intermittency, i.e. laminar and turbulent regions co-exist. Fully turbulent here specifies that prior to the first pressure tap ($484D$ from the inlet) the entire flow is in disordered motion and that the friction factor follows the Blasius friction scaling for Newtonian turbulence. The mixing chamber, which is a short $2D$ wide section of pipe, contains a resistance temperature detector (RTD) element and a polymer-injection port which connects to a pair of phase-offset peristaltic pumps through an air filled flow pulsation dampener. A second RTD probe was further added to the flexible piping between the pipe exit and the needle valve. A flexible silicone tube separating the mixing chamber from the glass pipe and a second pulsation dampener added to the inlet of the peristaltic pumps act in combination with the other dampener to insure minimal vibrations and near zero flow pulsations as a result of polymer injection. The injected polymers are in the form of a $600\text{ppm}$ solution of polyacrylamide (PAM, molecular weight 5-6 million, Polysciences Europe GmbH, lot number: 685910) dissolved in water, through a minimal shearing mixing process, over several days and kept at 296K. Note that while the molecular weight is only slightly larger than that of polymers used by Samanta et al. (2013), the present batch was considerably more efficient at drag reduction. Partially this may be attributed to the changed mixing procedure, in contrast to Samanta et al. (2013) vigorous stirring was avoided in the present study in favor of the rotating drum technique which minimized degradation during the mixing phase.

In the current study, the concentration of polymers in the pipe was adjusted by varying the injection rate of the concentrated polymer solution. The mixing chamber and the long development section of the pipe ensures homogeneous mixing of the concentrated polymer solution and the incoming water. The mixing efficiency of the polymers prior to the first pressure tap was tested by dyeing the concentrated solution and observing a homogeneous dye distribution far upstream of the tap. The measurement section begins at the first pressure tap and extends $132D$ in length ending at a second pressure tap for the differential pressure transducer dedicated to ascertaining the friction factor. A second differential pressure transducer was employed to measure pressure fluctuations along a $5D$ length of pipe. Laser Doppler velocimetry (LDV) and particle image velocimetry (PIV) measurements were performed through a water filled transparent box around $87D$ from the first pressure tap using a section of pipe which had a refractive index similar to that of water. PIV  measurements focused on the radial-axial cross section of the pipe. The PIV measurement window was $3D$ in length and images are mostly taken at a frequency of $\approx70\text{Hz}$. The high time resolution allows us to reconstruct larger flow structures by stitching several images together while taking the structure’s convection speed into account ($\text{i.e.}$ applying Taylor’s frozen turbulence hypothesis). Fluid exiting the pipe was collected and continuously weighed in order to assess the instantaneous mass flow rate, after which it was discarded.

Two alternative procedures were used to validate that the conclusions of this study were not dependent on the polymer injection method detailed above. In the first, premixed set concentrations of dilute polymer solutions were tested in the same setup by placing them in the elevated water reservoir and not using any injection. In the second, polymers were injected far downstream of the entrance and disturbance pin in order to observer whether polymer injection could fully suppress Newtonian turbulence after it has fully developed. The premixed solution proved to be move effective at suppressing Newtonian turbulence and reducing drag ($\text{i.e.}$ lower concentrations were needed to achieve the same effects). The downstream injections proved as effective as injections prior to the pipe entrance with a minor quantitative variation in terms of polymer concentration. Results from these alternative procedures will not be discussed in this paper apart from noting that overall the qualitative conclusions were consistent with what is presented in the main work.

\end{document}